\documentclass[twocolumn,showpacs,preprintnumbers,amsmath,amssymb,prl]{revtex4}
\usepackage{graphicx}
\usepackage{bm}
\usepackage{dcolumn}

\def\noi{\noindent}
\def\bc{\begin{center}}
\def\ec{\end{center}}
 \newcommand{\bea}{\begin{equation}}
 \newcommand{\eea}{\end{equation}\noi}
 \newcommand{\ber}{\begin{eqnarray}}
 \newcommand{\eer}{\end{eqnarray}\noi}
\begin{document}
\title{Resonance and limit cycle in a noise driven Lorenz model}
\author{Himadri   S.   Samanta}\email{h.s.samanta@sheffield.ac.uk}
\affiliation{Department of Physics and Astronomy,
University of Sheffield, Sheffield, S3 7RH, UK}
\author{J.   K. Bhattacharjee}\email{jkb@bose.res.in}
\affiliation{ S.N.Bose National Centre For Basic Sciences, JD-Block,Sector-III, Salt Lake City, Kolkata-700098, India}
\date{\today}
\begin{abstract}
The effect of an external noise on the Lorenz model is
investigated near the onset
        of convection and near the Hopf bifurcation. We show the
existence of a diverging time scale
        near the onset of convection and a resonance near the Hopf
bifurcation. Our
        calculation provides an understanding of the noise induced
stabilization of the limit
       cycle that had been observed numerically.
\end{abstract}

\maketitle


The noise induced phenomena have an intense interest in several disciplines ranging from 
physics, to chemistry, and to biology \cite{9}. Physical systems are usually not isolated from
their environment, and the environmental influences on the system often appear as fluctuations.
It is clear that the fluctuations are an integral part of the evolution of physical, chemical 
and biological systems and must be understood if we are to accurately and quantitatively 
describe the world around us, particularly at the small scale.
In the past few years, it has 
become clear that fluctuations can actually be used constructively, by us or by nature, to produce
organised behavior that is not possible in the absence of noise. Examples where noise leads to 
organised behavior include stochastic resonance \cite{1,2}, noise induced phase transitions \cite{3,4}, noise induced
pattern formation \cite{5,6}, and noise induced transport \cite{8,9}. The constructive role of noise is only possible 
in the non-linear non-equilibrium systems and is entirely the result of the intricate interplay
of noise and non-linearity away from equilibrium.
 
The effect of noise on dynamical systems include noise induced hopping between multiple stable attractors\cite{10,11}
and noise induced stabilization of the Lorenz attractor \cite{7,omar} near the threshold of its formation. 
We will focus on the second aspect here and as an example of the effect of noise on hydrodynamic instability. 
We consider the Lorenz model with an external noise source. Zippelius et. al.\cite{luke} first studied 
the correlations in this model. We will supplement their work by focusing specially on the situations 
where the control parameter is in the vicinity of an instability. This will lead to a noise induced 
resonance near the Hopf bifurcation point. We then focus on the effect that the noise can have on the limit cycle 
beyond the Hopf bifurcation point. Ordinarily the limit cycle is unstable, the numerical work of Gao et. al. \cite{7}
shows it can be stabilized by noise and here we explicitly show how such a stabilization is possible
by drawing on the technique of Mclaughlin and Martin\cite{martin}.

We study the statistical properties of the following noise driven Lorenz model.

\ber\label{1}
\dot{x}&=&\sigma(y-x) +\eta_{1}(t)\\ \nonumber
\dot{y}&=& r x -y-x z+\eta_{2}(t)\\ \nonumber
\dot{z}&=&-b z+x y +\eta_{3}(t)
\eer

where, $\sigma$ is the Prandtl number, $b$ is a geometric factor, and the Rayleigh number 
$r$ is the control parameter in this model. The parameter $\sigma=10$ and $b=8/3$ are 
held fixed at their standard values, as was originally done by Lorenz. $\eta_{i}(t)$ is 
an external white noise source with zero mean and $<\eta_{i}(t)\eta_{j}(t^{'})>=\delta_{ij}
2 D \delta(t-t^{'})$, $i,j=1,2,3$. $D$ is the noise strength. $D=0$ describes the unforced
Lorenz model. When the noise is absent the Lorenz system \cite{lorenz,row} shows remarkable change in behavior 
depending on the control parameter $r$. In the conduction range, $r<1$, the trivial steady 
state solution is stable and loses stability to the other two describing steady convection
through a bifurcation at $r=1$. Thus, for $r>1$, there is a pair of stable fixed points,
$(\pm \sqrt{b(r-1)}, \pm \sqrt{b(r-1), r-1})$ and these in turn lose stability at $r=r_{T}
=\sigma(\sigma+b+3)/(\sigma-b-1)=24.74$ through a Hopf-bifurcation. For $r>r_{T}$, no stable
steady state solution exists and the system has a strange attractor. Depending on initial 
conditions, the solution may settle down on any of these three attractors. The trajectories 
are non-periodic and wander around in the vicinity of strange set of attracting points for
$r>r_{T}$ and turbulence sets in.

Now we study the effect of noise on the Lorenz model in the conduction, convection
and turbulent regime. We calculate the time dependent correlation functions and compare with
the behavior of the unforced model.


In frequency space equation (\ref{1}) reads

\ber\label{2}
&&(-i \omega +\sigma) x(\omega)-\sigma y(\omega)=\eta_{1}(\omega)\\ \nonumber
&& -r x(\omega)+(-i\omega+1)y(\omega)+\sum_{\omega_{1}}x(\omega_{1})
z(\omega-\omega_{1})=\eta_{2}(\omega) \\ \nonumber
&& (-i\omega+b)z(\omega)-\sum_{\omega_{2}}x(\omega_{2})y(\omega-\omega_{2})=\eta_{3}(\omega)
\eer

Linearizing around the steady state $x=y=z=0$ i.e. neglecting the non-linear terms in (\ref{2}),
we can calculate the correlation function by solving the linearized equation. The correlation 
of the $x$ variable reads

\ber\label{3}
 C_{xx}(\omega)&=& <x(\omega)x(-\omega)> \\ \nonumber
&=& \frac{(\omega^{2}+1)+\sigma^{2}}{([\omega^{2}+(r-1)\sigma]^2+(\sigma+1)^2 \omega^2)}
\eer

The time dependent correlation function $C_{xx}(t)$ can be written as
\bea\label{4}
C_{xx}(t)=\int_{-\infty}^{\infty}d \omega e^{-i\omega t} C_{xx}(\omega)
\eea

and is the sum of two exponential function with two inverse relaxation times which are 
independent of the strength $D$ of the fluctuating force. Clearly, the actual relaxation
time of the non-linear system (\ref{1}) does depend on the strength of the fluctuating 
force. The correlation function of $x$ gets damped out faster with increasing noise strength $D$.
The correlation functions $C_{yy}(t)$ and $C_{zz}(t)$ follow the same behavior in the regime $r<1$.

Near, $r=1$, the correlation time goes as
\bea\label{5}
\tau \propto \frac{1}{r-1}
\eea

i.e. the relaxation time becomes infinitely big as $r=1$ is approached- a sign of critical slowing down.

Now, for $r > 1$, Fourier analysis of the Lorenz model breaks down in its present form.
This is very similar to what happens when one enters a symmetry breaking phase in critical phenomena.
Accordingly we need to go to the shifted variable $u_1 , u_2, u_3$ defined as
$u_1=x-x_0$, $u_2=y-y_0$ and $u_3=z-z_0$.

The Lorenz equation takes the following form

\ber\label{8}
&&(-i \omega +\sigma)u_1(\omega)-\sigma u_2(\omega)=\eta_{1}(\omega)\\ \nonumber
&&-u_1(\omega)+(-i \omega +1)u_2(\omega)+x_{0} u_3(\omega)=\eta_{2}(\omega) \\ \nonumber
&&-y_{0}u_1(\omega)-x_{0}u_2(\omega)+(-i \omega+b)u_3(\omega)=\eta_{3}(\omega)
\eer

Now, the correlation function takes the following form
\ber\label{9}
&& <u_{1}(\omega)u_{1}(-\omega)>= \\ \nonumber
&& \frac{(-\omega^{2}+br)^{2}+(b+1)^2 \omega^{2}+\sigma^{2}(b^2+\omega^2)+\sigma^2 b (r-1)}
{\omega^2 [\omega^2 -b(r+\sigma)]^2 +[\omega^2 (\sigma+1+b)-2\sigma b (r-1)]^2}
\eer

The time dependent correlation function can be calculated from
 $C_{u_{1}u_{1}}(t)=\int_{-\infty}^{\infty}d \omega e^{-i\omega t} C_{u_{1}u_{1}}(\omega)$.
Their time dependence is determined by the three poles in the complex frequency 
plane. One is purely imaginary reflecting the exponential decay in the correlation function,
the other two have finite real parts of the opposite sign, reflecting spiral motion
around one of the attractors ($\pm x_0, \pm y_0, z_0$). This gives rise to the oscillatory 
behavior of $C_{u_{1}u_{1}}(t)$ superimposed upon the exponential decay of the correlations 
caused by the crossing between stable fixed points. In the absence of random force, 
the system is attracted in general to one of the stable fixed points ($\pm x_0, \pm y_0, z_0$),
depending on its initial condition. If the random force is applied, the trajectories are 
no longer confined to one of the steady state points. The exponential decay shows the motion 
from one fixed point to the other. The oscillatory motion slows down the decay of correlation function
i.e. the memory effect of the initial state.

At the Hopf-bifurcation point, i.e. $r=\frac{\sigma(b+\sigma+3)}{\sigma-b-1}$, the correlation function
takes the following form
 
\ber\label{10}
&& <u_{1}(\omega)u_{1}(-\omega)>= \\ \nonumber
&& \frac{(-\omega^{2}+br)^{2}+(b+1)^2 \omega^{2}+\sigma^{2}(b^2+\omega^2)+\sigma^2 b (r-1)}
{[\omega^2 -\omega_{0}^2)]^2 [\omega^2 + (\sigma+1+b)^2)]}
\eer

where, $\omega_0=b(\sigma+\frac{\sigma(b+\sigma+3)}{\sigma-b-1})$. Now, the real time correlation function
$C_{u_{1}u_{1}}(t)=\int_{-\infty}^{\infty}d \omega e^{-i\omega t} C_{u_{1}u_{1}}(\omega)$ exists as a principle value.
Hence, the real time correlation function goes as

\bea\label{11}
C_{u_{1}u_{1}}(t)\propto Re \ t e^{i \omega_{0}t}
\eea

This is a consequence of the fluctuating force. It is similar in appearance to the resonance in a 
simple harmonic oscillator subjected to a sinusoidal force.

Approach to this resonance is of the following

\bea\label{13}
<u_{1}u_{1}>=\lim_{\epsilon->0}\frac{1}{\epsilon} e^{i\omega t}
\eea

when, $r=\frac{\sigma(b+\sigma+3)}{\sigma-b-1}-\epsilon$.

 In the absence of noise there is no periodic
state above the Hopf bifurcation point. This is because the
limit cycle is unstable.In this case however,a periodic state
was observed when the stochastic force was turned on. 
We will try to understand this on the basis of perturbation theory in terms of small 
noise strength $\epsilon$. We return to Eq.(8). Writing $r=r_0 +\Delta r \epsilon$,
$x_0 =x_{00}+\tilde{\epsilon}$, where $x_{00}=\sqrt{b(r-1)}$, $\tilde{\epsilon}=\frac{\epsilon \Delta r}{2}
\sqrt{\frac{b}{r-1}}$, we find in real time

\bea\label{l1}
L \left(\begin{array}{c} u_1 \\ u_2 \\u_3 \end{array}\right)
=\left(\begin{array}{c}\eta_1 \\\eta_2 \\\eta_3   \end{array}\right)+
\left(\begin{array}{c}0\\-u_1 u_3\\u_1 u_2  \end{array}\right)+
\tilde{\epsilon}\left(\begin{array}{c}0\\-u_3\\u_1 +u_2   \end{array}\right)
\eea

where the operator $L$ is given by

\bea\label{l2}
L=\left(\begin{array}{c}\begin{array}{cc}\begin{array}{ccc}\frac{\partial}{\partial t}+\sigma &-\sigma & 0 \\
-1 & \frac{\partial}{\partial t}+1 & x_{00} \\ 
-x_{00} & -x_{00} & \frac{\partial}{\partial t}+b \end{array}\end{array}\end{array}\right)
\eea

Consistent ordering in powers of $\epsilon$, leads to the expansion 
\bea\label{l3}
\left(\begin{array}{c}u_1 \\u_2\\u_3\end{array}\right)=
\epsilon^{1/2}\left(\begin{array}{c} u_{10}\\u_{20}\\u_{30} \end{array}\right)
+\epsilon \left(\begin{array}{c}u_{11}\\u_{21}\\u_{31}   \end{array}\right)
+\epsilon^{3/2}\left(\begin{array}{c}u_{12}\\u_{22}\\u_{32}  \end{array}\right)
\eea

The zeroth order solution is found from 

\bea\label{l4}
L \left(\begin{array}{c}u_{10}\\u_{20}\\u_{30}  \end{array}\right)=
\left(\begin{array}{c}\eta_{1}\\\eta_{2}\\\eta_{3}  \end{array}\right)
\eea

as

\bea\label{l5}
u_{i0}= Re A_{i} e^{i w t} +\int G_{ij}(t-t')\eta_{j}(t')dt''
\eea
\begin{figure}[t]
\begin{center}
\includegraphics[scale=0.9]{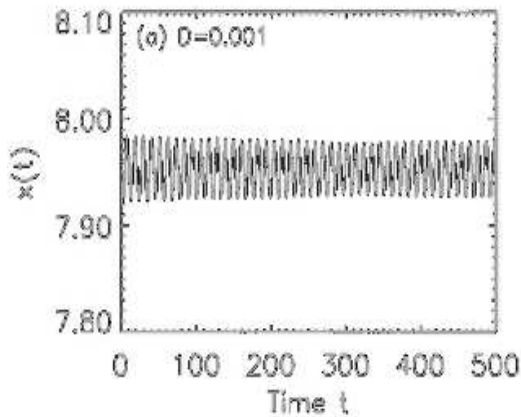}
\caption{\small{ Time series for the Lorenz equations with
r=24.72, D= 0.001 as obtained by Gao et. al.
}}
\end{center}
\end{figure}

where, $A_1 =A$, $A_2=\frac{A}{\sigma}(\sigma-iw)$, $A_3=\frac{A}{x_{00}}[1-\frac{(1-i w_0)(\sigma-i w_0)}{\sigma}]$
and $G_{ij}$ are found in frequency space from the particular integral of Eq.(\ref{l4}).
At $\mathcal{O}(\epsilon)$, we have

\bea\label{l6}
L \left(\begin{array}{c}u_{11} \\ u_{21}\\u_{31}\end{array}\right)
=\left(\begin{array}{c}0\\ -u_{10} u_{30} \\ u_{10}u_{20}
\end{array}\right)
\eea

and the driving term in the above equation has contributions which are;\\
(i) a time independent term,\\
(ii) a time periodic term with frequency $2 w_{0}$,\\
(iii) a term which is a product of a periodic term of frequency $w_{0}$ and 
a stochastic term $\eta$ and \\
(iv) a term which is a product of two stochastic terms.

The solutions for $u_{i1}$ is a contribution of these driving terms with suitable coefficients.

Having obtained $u_{i1}$, we can now go to $\mathcal{O}(\epsilon^{3/2})$ and we have
\begin{widetext}
\bea\label{l7}
L\left(\begin{array}{c}u_{12}\\u_{22}\\u_{32}\end{array}\right)=
\left(\begin{array}{c} 0\\ -u_{11}u_{30}-u_{10}u_{31}\\ u_{10}u_{21}+u_{11}u_{20}\end{array}\right)
+\frac{\Delta r}{2}\sqrt{\frac{b}{r-1}}\left(\begin{array}{c}0\\-u_{30}\\u_{10}+u_{20}\end{array}\right)
\eea
\end{widetext}
The driving terms on the right hand side need to be analysed. The second term is time periodic 
with frequency $w_{0}$ and has an amplitude proportional to $(\Delta r)A$.
In the first term, we need to analyse a typical term ($u_{11}u_{30}$ e.g.) to understand the different time dependent 
components of the drive. The term $u_{30}$ is periodic with frequency $w_{0}$. 
Two of the components of $u_{11}$ as shown above are constant in time and periodic with frequency $2 w_{0}$.
The product $u_{30}u_{11}$ then has terms with frequency $w_{0}$ and the amplitude of these terms 
 are proportional to $A^3$. The operator $L$ has a zero mode at frequency $w_{0}$, consequently in order to allow
a finite solution of Eq.(\ref{l7}), the right hand side has to be orthogonal to the left eigenvector of $L$
with zero eigenvalue. In the absence of the stochastic drive, this is the entire calculation and the result is that 
$\Delta r =-\beta^2 A^2$, where $\beta$ is a constant. Thus there is no real value of $A$ if $\Delta r$ is positive
and the limit cycle of the Lorenz model cannot be seen.

In the presence of stochastic term, things change because $u_{i0}$ (Eq.(\ref{l4})) has a term 
proportional to $\eta$ and $u_{i1}$ has the two terms which are listed in (ii) and (iv) above.
If we consider $u_{i0}u_{j1}$ then there are
two terms whose solution is $A \eta \eta e^{iw_{0}t}$.
Since we are in the presence of stochastic terms we can only talk about averages over $\eta$.
Consequently, we first need to average Eq.(\ref{l7}) over $\eta$ and only terms involving product 
of two $\eta$'s will survive. After the averaging the term with structure $A \eta \eta e^{iw_{0}t}$
will acquire the structure $A  e^{iw_{0}t}$ and will be a part of the dangerous term on the right 
hand side. The orthogonality condition now leads to an equation $\Delta r + N=-\beta^2 A^2$, 
where $N$ is the extra contribution coming from the stochastic term. In this particular case $N$
 is negative and hence for $\Delta r< N$, we can see the limit cycle stabilized and this is the mechanism 
which allows the time series to be periodic in a region of $r$, where there is no stable limit cycle.

In closing, we have studied the statistical properties of noise driven Lorenz model. We have seen critical slowing down 
at stationary bifurcation point. At $r>1$, the Fourier analysis breaks down due to lack 
of time translational invariance in present form of Lorenz model. 
An interesting resonance appears at the Hopf-bifurcation point,
which shows the the noise induced stability. We also analysed the noise induced stability 
of limit cycle in the Lorenz model.


\begin{thebibliography}{99}
\bibitem{9}P. Reimann, Phys. Rep. {\bf361}, 57 (2002).
\bibitem{1} R. Benzi, A. Sutera, and A. Vulpiani, J. Phys. A {\bf 14}, L453, (1981).
\bibitem{2}L. Gammaitoni, P. Hanggi, P. Jung, and F. Marchesoni, Rev. Mod. Phys. {\bf 70}, 223 (1998).
\bibitem{3} C. Van den Broeck, J. M. R. Parrondo, and R. Toral, Phys. Rev. Lett. {\bf 73}, 3395 (1994);
C. Van den Broeck, J. M. R. Parrondo, J. Armero, and A. Hernández-Machado, Phys. Rev. E {\bf 49}, 2639 (1994).
\bibitem{4} G. Grinstein, M. A. Muñoz, and Y. Tu, Phys. Rev. Lett. {\bf 76}, 4376 (1996). 
\bibitem{5} J. Buceta, M. Ibañes, J. M. Sancho, and Katja Lindenberg,
Phys. Rev. E {\bf 67},0 021113 (2003).
\bibitem{6} M.C. Cross and P.C. Hohenberg, Rev. Mod. Phys. {\bf 65}, 851 (1993).
\bibitem{7} J. B. Gao, Wen-wen Tung and Nageswara Rao, Phys. Rev. Lett. {\bf89}, 254101 (2002).
\bibitem{8}F. Julicher, A. Ajdari and J. Prost Rev. Mod. Phys. {\bf69}, 1269 (1997). 
\bibitem{10}R. L. Kautz, J. Appl. Phys. {\bf58}, 424 (1985).
\bibitem{11}F. Arecchi, R. Badii, and A. Politi, Phys. Rev. A {\bf32}, 402 (1985).
\bibitem{luke}  Annette Zippelius and Manfred L$\ddot{u}$cke, J. Stat. Phys. {\bf 24}, 345 (1981)
\bibitem{martin} McLaughlin, J. B. and Martin, P. C., Phys. Rev. A {\bf 12}, 186 (1973).
\bibitem{lorenz}Lorenz, E. N., J. Atmos. Sci. {\bf 20}, 130 (1962).
\bibitem{row} Rowlands, G., J. Phys. A {\bf 16}, 585 (1983)
\bibitem{omar}Omar Osenda, Carlos B. Briozzo, and Manuel O. Caceres, Phys. Rev. E {\bf 55}, R3824 (1997).
\end{thebibliography}
\end{document}